\documentclass[%
 reprint,
 amsmath,amssymb,
 aps,
]{revtex4-1}

\usepackage{graphicx}            
\usepackage{dcolumn}             
\usepackage{bm}                  
\usepackage{hyperref}            
\usepackage[mathlines]{lineno}   
\usepackage{url}								 
\usepackage{comment}             
\usepackage{color}               

\graphicspath{{./Figures/}}

\begin{document}

\preprint{APS/123-QED}

\title{Design and operation of a prototype interaction point beam collision feedback system for the International Linear Collider}

\author{R.\,J.~Apsimon}
  \altaffiliation{Present address: Cockcroft Institute, University of Lancaster, UK.} 
\author{D.\,R.~Bett}
\author{N.~Blaskovic~Kraljevic}
  \altaffiliation{Present address: CERN, Geneva, Switzerland; \\ \href{mailto:neven.blaskovic.kraljevic@cern.ch}{neven.blaskovic.kraljevic@cern.ch}.}
\author{R.\,M.~Bodenstein}
\author{T.~Bromwich}
\author{P.\,N.~Burrows}
\author{G.\,B.~Christian}
  \altaffiliation{Present address: Diamond Light Source, Harwell, UK.}
\author{B.\,D.~Constance}
\author{M.\,R.~Davis}
\author{C.~Perry}
\author{R.~Ramjiawan}

 \affiliation{John Adams Institute for Accelerator Science at University of Oxford, Denys Wilkinson Building, Keble Road, Oxford OX1 3RH, United Kingdom}

\date{\today}

\begin{abstract}

A high-resolution, intra-train position feedback system has been developed to achieve and maintain collisions at the proposed future electron-positron International Linear Collider (ILC). A prototype has been commissioned and tested with beam in the extraction line of the Accelerator Test Facility at the High Energy Accelerator Research Organization (KEK) in Japan. It consists of a stripline beam position monitor (BPM) with analogue signal-processing electronics, a custom digital board to perform the feedback calculation and a stripline kicker driven by a high-current amplifier. The closed-loop feedback latency is 148~ns. For a three-bunch train with 154~ns bunch spacing, the feedback system has been used to stabilize the third bunch to 450~nm. The kicker response is linear, and the feedback performance is maintained, over a correction range of over $\pm 60$~$\mu$m. The propagation of the correction has been confirmed by using an independent stripline BPM located downstream of the feedback system. The system has been demonstrated to meet the BPM resolution, beam kick and latency requirements for the ILC.



\end{abstract}

														

\maketitle


\section{\label{sec:Intro}Introduction}

\begin{figure*}
	\centering
	\setlength\fboxsep{0pt}
	\includegraphics[width=\textwidth]{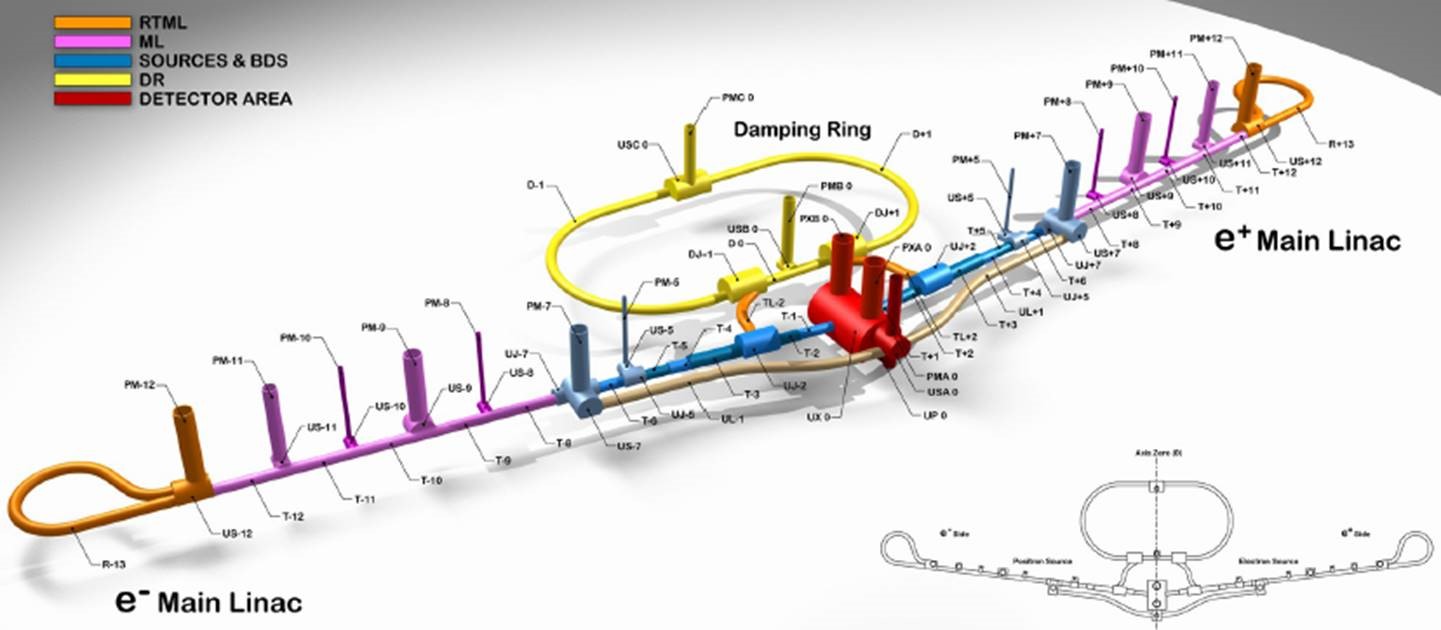}
	\caption{Schematic layout of the ILC \cite{ILCTDRvol3}.}
	\label{fig:ILCSchematic}
\end{figure*}

The International Linear Collider (ILC)~\cite{ILCTDRvol3} is a proposed high-luminosity electron-positron collider (Fig.~\ref{fig:ILCSchematic}) with a baseline center of mass (c.m.) energy of 500~GeV and options for operating within the c.m. energy range between 250 and 1000~GeV. The 500~GeV baseline design luminosity of $1.8 \times 10^{34}$~cm$^{-2}$~s$^{-1}$ requires long trains of 1312 particle bunches and colliding beams focused at the interaction point (IP) to $\sim 6$~nm (vertical) and $\sim 500$~nm (horizontal). The design parameters for the 250, 500 and 1000~GeV machines are shown in Table~\ref{tab:ILCParameters}. In order to compensate for residual vibration-induced jitter from the final focus magnets at frequencies near and above the bunch-train repetition frequency of 5~Hz, a fast, intra-train IP beam position feedback system is required to maintain bunch collisions over the course of each train \cite{Schulte99}. The short inter-bunch time separation of 554~ns demands that such a feedback system has a low latency so as to allow for the possibility of bunch-by-bunch corrections. Here we present the design, commissioning and operation of a prototype IP feedback system that meets the ILC requirements. 

\begin{table}
\caption{Parameters for the ILC 250~GeV and 500~GeV baseline designs and for an energy upgrade to 1~TeV \cite{ILCTDRvol1}.}
\label{tab:ILCParameters}
\centering
\begin{tabular}{ccccc}
																																												  \hline \hline 
		Center of mass energy   & GeV                                 & 250  & 500  & 1000 \\ \hline 
		Collision rate          & Hz                                  & 5    & 5    & 4    \\
		Number of bunches       &                                     & 1312 & 1312 & 2450 \\
		Bunch population        & $\times 10^{10}$                    & 2.0  & 2.0  & 1.74 \\
		Bunch separation        & ns                                  & 554  & 554  & 366 \\
		IP horizontal beam size & nm                                  & 729  & 474  & 335 \\
		IP vertical beam size   & nm                                  & 7.7  & 5.9  & 2.7 \\
		Luminosity              & $\times 10^{34}$~cm$^{-2}$~s$^{-1}$ & 0.75 & 1.8  & 4.9 \\ \hline \hline 


\end{tabular}
\end{table}

Since the vertical beam size at the IP is roughly 100 times smaller than the horizontal beam size, the vertical axis is most sensitive to relative beam-beam misalignments and hence we describe a system for making beam trajectory corrections in the vertical plane. A corresponding system could operate in the horizontal plane.

\begin{figure}
	\centering
  \includegraphics[width=\columnwidth]{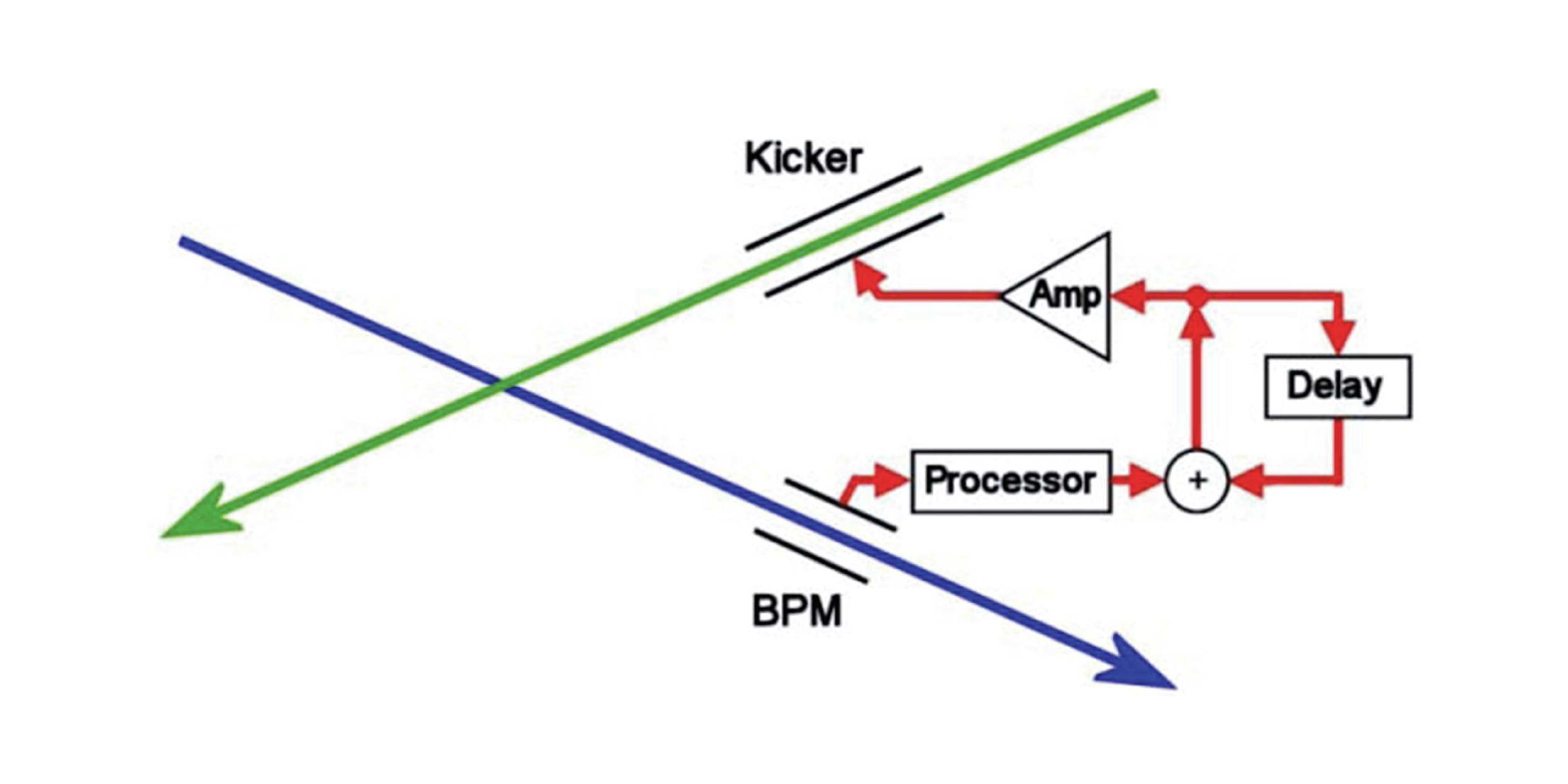}
	\caption{Functional schematic of the intra-train IP beam feedback system \cite{Smith01}.}
	\label{fig:IPFeedbackPrototype}
\end{figure}

\begin{figure}
	\centering
  \includegraphics[width=\columnwidth]{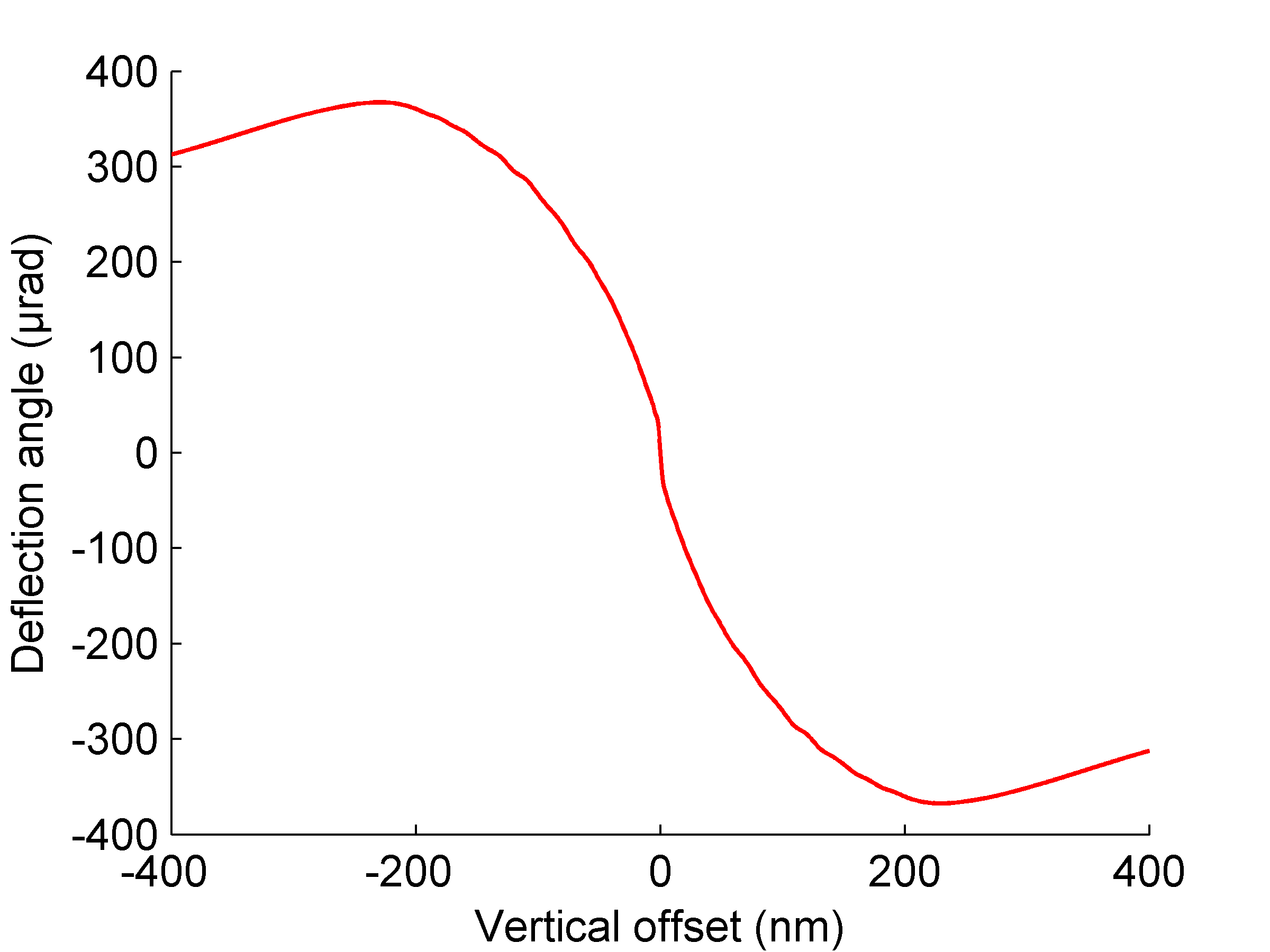}
	\caption{Outgoing beam deflection angle versus incoming relative beam position offset at the IP for the ILC baseline design at 500~GeV c.m. energy.}
	\label{fig:AngleVsOffset}
\end{figure}

\begin{figure*}
	\centering
	\setlength\fboxsep{0pt}
	\includegraphics[width=\textwidth]{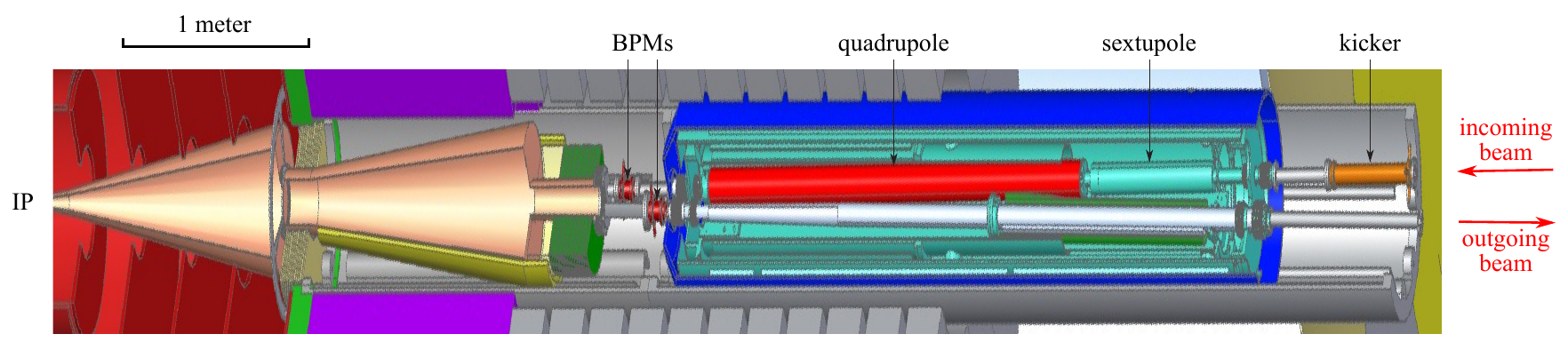}
	\caption{Engineering schematic \cite{MarkiewiczPC} of the IP region showing the location of the kicker on the incoming beamline and the feedback BPM on the outgoing beamline. A vertical-to-horizontal aspect ratio of 3:1 has been used in this figure.}
	\label{fig:IPFeedbackEngineering}
\end{figure*}

A schematic of the proposed intra-train IP feedback system for correction of the relative vertical beam misalignment is shown in Fig.~\ref{fig:IPFeedbackPrototype} for the case in which the two beams cross with a small horizontal angle; the ILC design incorporates a crossing angle of 14~mrad. The system relies on the strong transverse electromagnetic kick experienced by each electron bunch in the field of the opposing positron bunch (and vice versa) when the two bunches arrive at the IP with a relative vertical offset \cite{Chen13}.  Beam simulations, performed using the tracking code Lucretia \cite{Tenenbaum05} and the beam-beam interaction code \textsc{guinea-pig} \cite{SchultePhD}, allow the deflection angle to be calculated as a function of the relative offset at the IP of the incoming bunches (Fig.~\ref{fig:AngleVsOffset}); the results presented here complement earlier simulations \cite{White06, Resta08, Resta09, Bodenstein17} performed using the tracking code \textsc{placet} \cite{DAmico01}. The ILC lattice has been used with the final focus length $L^*$ updated to 4.1~m. The capture range of the ILC IP intra-train feedback system has been specified to be $\pm 200$~nm relative beam offset \cite{Parker09}. It can be seen from Fig.~\ref{fig:AngleVsOffset} that within this range the angular deflection imparted to the outgoing bunches varies with the relative bunch offset, and spans the range within roughly $\pm350$~$\mu$rad. Such a large outgoing beam deflection angle causes beam position displacements of up to 1400~$\mu$m in a beam position monitor (BPM) placed $\sim 4$~m downstream of the IP as in the ILC design~\cite{ILCTDRvol3}. The BPM signals are processed to derive a correction signal which is amplified and used to drive a kicker, located $\sim 8$~m upstream of the IP, on the other incoming beamline (Fig.~\ref{fig:IPFeedbackPrototype}). An engineering implementation is shown in Fig.~\ref{fig:IPFeedbackEngineering}.


If the total system latency is shorter than the inter-bunch separation, the trajectory of each successive incoming bunch can be corrected. The measurement and correction are performed on opposing beams so as to reduce the signal propagation time between BPM and kicker \cite{Hendrickson99}, and the beamline components are placed as close as possible to the IP in order to minimize the delay due to the total beam flight time from the kicker to the IP and from the IP to the BPM. Since the system acts on each successive bunch crossing a delay loop (Fig.~\ref{fig:IPFeedbackPrototype}) is required, which constitutes a memory of the sum of preceding corrections and maintains the correction for subsequent bunches.

\begin{figure}
	\centering
  \includegraphics[width=\columnwidth]{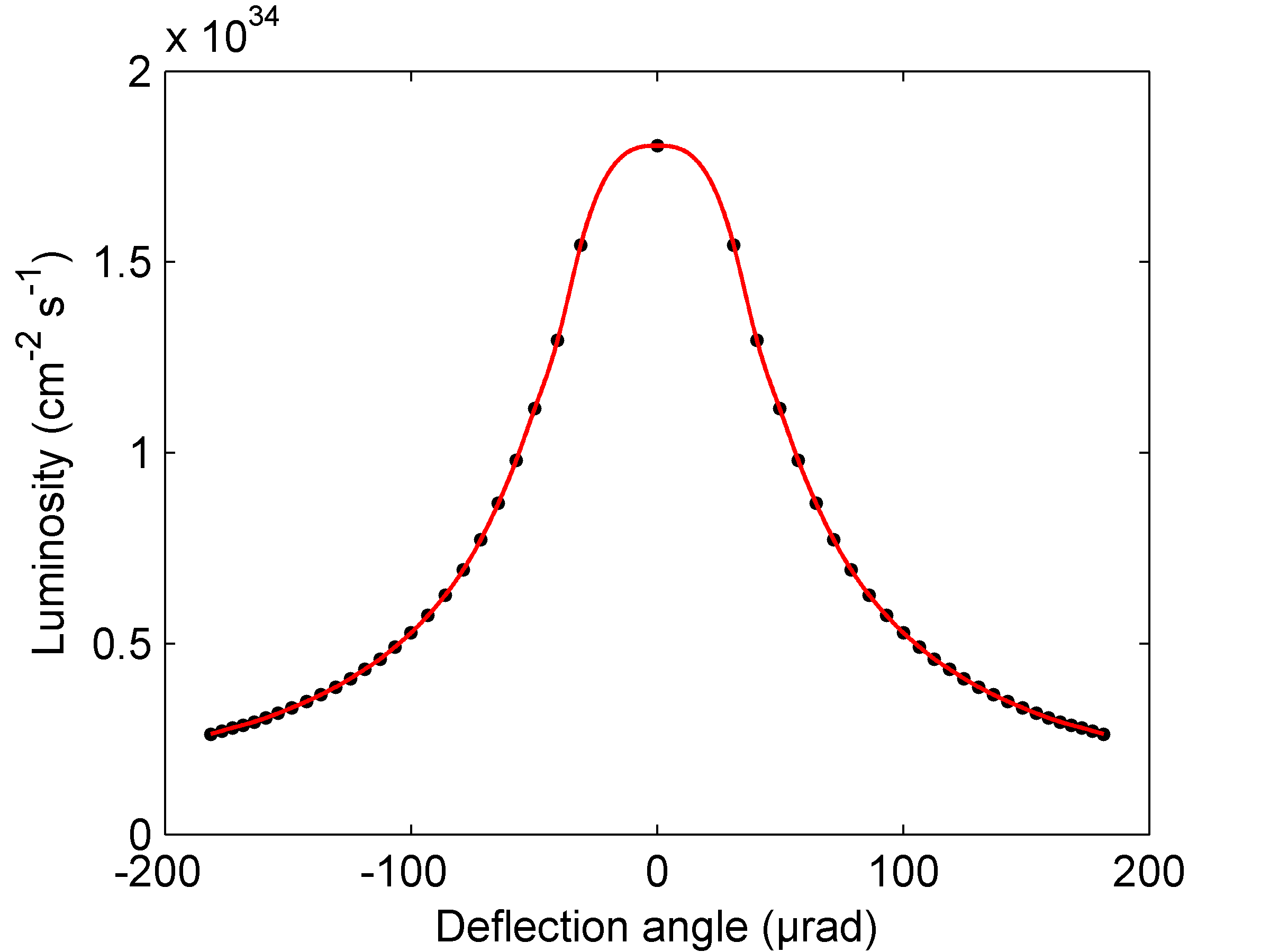}
	\caption{Luminosity versus beam-beam deflection angle. The red line is a cubic spline interpolation.}
	\label{fig:LuminosityVsAngle}
\end{figure}

Using the beam simulations described above, the luminosity has been calculated as a function of the beam-beam deflection angle (Fig.~\ref{fig:LuminosityVsAngle}). A 1\% degradation of the peak luminosity corresponds to a deflection angle of 13~$\mu$rad, which would be measured as a $\sim 50$~$\mu$m deflection at the BPM. Hence, a micron-level resolution for the feedback BPM is more than adequate to enable precise luminosity optimization and the requirement of a $\pm 1400$~$\mu$m linear range will handle deflection angles of up to $\pm 350$~$\mu$rad. The kicker is required to have sufficient drive to correct an IP relative bunch offset of up to $\pm 200$~nm which, for the ILC final focus magnets, sets a $\sim \pm 60$~nrad kick range requirement for a 250~GeV beam.


\section{\label{sec:Setup}Experimental setup}

\begin{figure}
  \centering
  \includegraphics[width=\columnwidth]{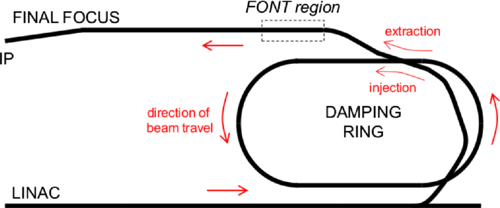}
	\caption{Layout of the ATF showing the location of the FONT system.}
	\label{fig:ATFSchematic}
\end{figure}

A prototype of such a feedback system has been developed by the Feedback on Nanosecond Timescales (FONT) group \cite{FONTurl} and has been installed, commissioned and tested at the Accelerator Test Facility (ATF) \cite{Hinode95} at KEK. The ATF (Fig.~\ref{fig:ATFSchematic}) is a 1.3~GeV electron test accelerator for the production of very low emittance electron beams as required for future linear electron-positron colliders. In 2008, as part of the ATF2 project~\cite{ATF2proposal1}, the beamline was upgraded and the extraction line was replaced with one incorporating an energy-scaled version of the compact beam focusing system designed for linear colliders \cite{Raimondi01}. The goals of the ATF2 Collaboration \cite{Bambade10} are to produce a 37~nm vertical beam spot size at the final focus point and to stabilize the vertical beam position to the nanometer level.

\begin{figure}
  \centering
  \includegraphics[width=\columnwidth]{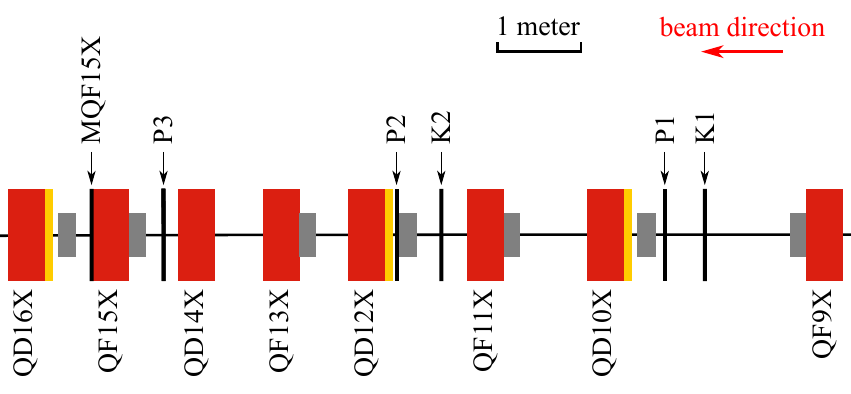}
	\caption{Layout of the stripline BPMs (P1, P2, P3 and MQF15X) and kickers (K1 and K2) used in the FONT system. Quadrupole magnets (`Q') are shown in red, skew quadrupoles in yellow and correctors in gray.}
	\label{fig:EXTFFSchematic}
\end{figure}

\begin{figure}
  \centering
  \includegraphics[width=\columnwidth]{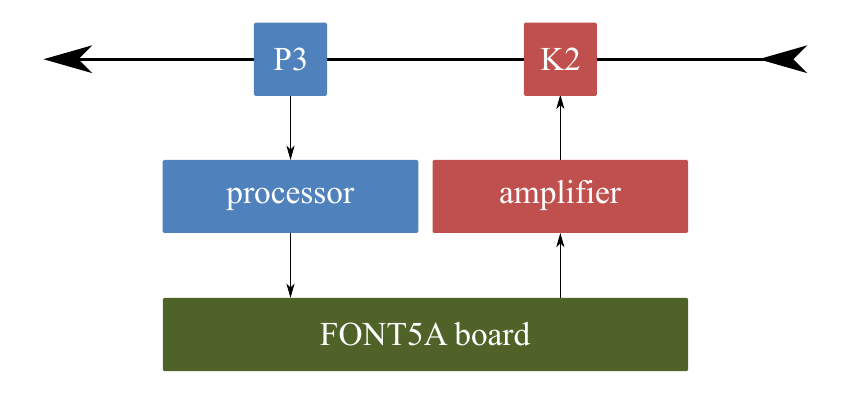}
	\caption{Block diagram of the single-loop feedback system using BPM P3 and kicker K2.}
	\label{fig:FeedbackSchematic}
\end{figure}

In order to address the ATF2 beam stabilization goals, the FONT group has developed a beam position stabilization system (`FONT5') \cite{Apsimon12} which is deployed in the upstream part of the ATF extraction line (Fig.~\ref{fig:ATFSchematic}). The full feedback system has been designed to stabilize both the beam position and angle in the vertical plane such that a fully-corrected beam can  propagate downstream into the ATF final focus line. For this purpose, the feedback system comprises the stripline BPMs P2 and P3, and the stripline kickers K1 and K2, whose beamline layout is shown in Fig.~\ref{fig:EXTFFSchematic}. The BPMs P1 and MQF15X are independent of the feedback loop, and are used as witnesses of the incoming and outgoing beam trajectories, respectively. Furthermore, in the context of the demonstration of an ILC-like IP position feedback system, the FONT5 system has been operated in `single-loop' mode using P3 to measure the vertical beam offset and K2 to correct it (Fig.~\ref{fig:FeedbackSchematic}). The hardware components are described below.

\subsection{\label{sec:StriplineBPM}Stripline BPM and processor}

\begin{figure}
  \centering
  \includegraphics[width=0.5\columnwidth]{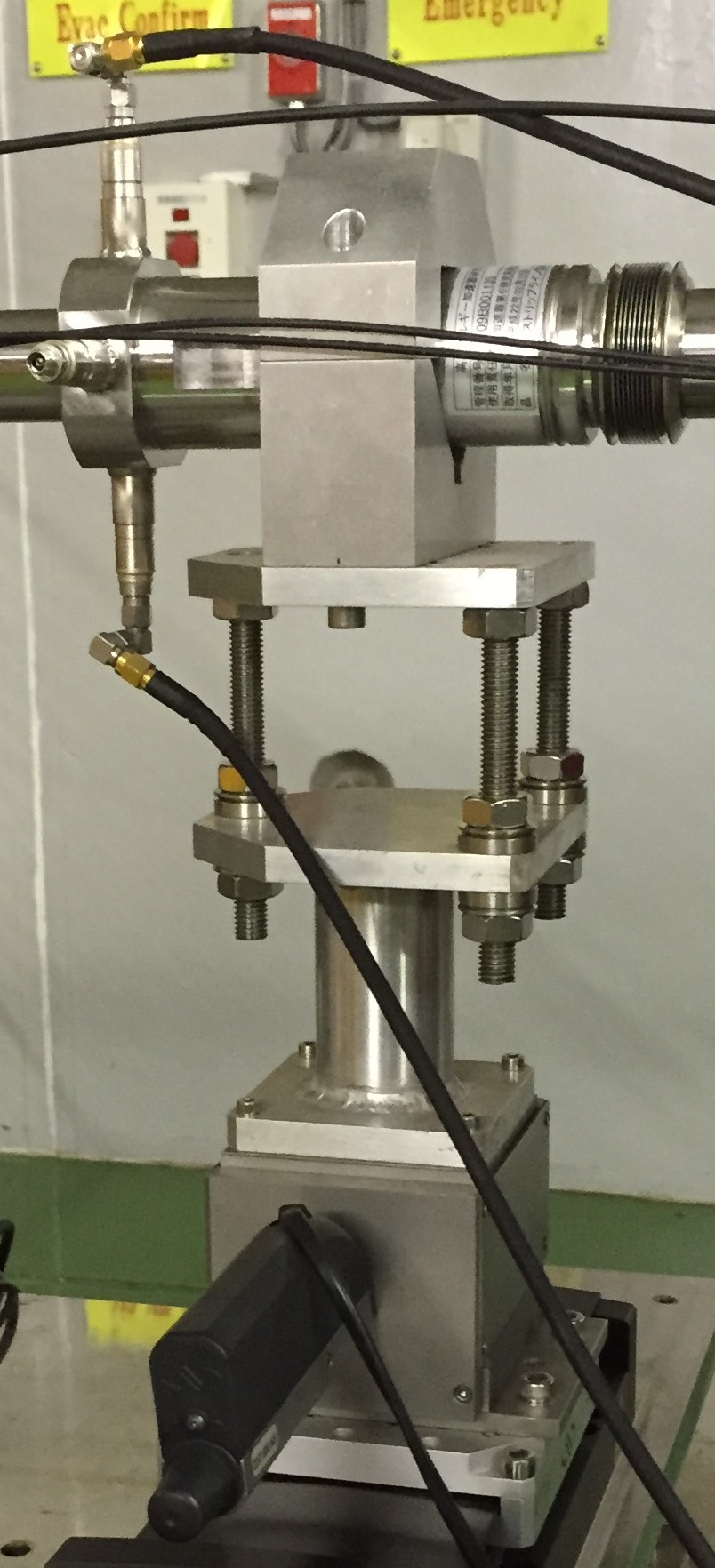}
	\caption{Photograph of the stripline BPM P3 and its mover in the ATF beamline.}
	\label{fig:StriplineBPM}
\end{figure}

The FONT stripline BPMs (Fig.~\ref{fig:StriplineBPM}) each consist of four 12-cm-long strips, arranged as two orthogonal diametrically opposed pairs separated by 23.9~mm~\cite{Apsimon15}. BPMs P1, P2 and P3 are each mounted on a M-MVN80 and M-ILS50CCL Newport mover system \cite{FausGolfePC} that can translate the BPM vertically and horizontally in the plane perpendicular to the beam, allowing the beam to be centered within each BPM aperture. BPM MQF15X, located 0.76~m downstream of P3, was not placed on a mover system.

The analogue signal processors have been developed specifically for high resolution and low latency. A single BPM processor can be used to process the beam position data in either the horizontal or vertical axis; from here on only the vertical plane is considered. The BPM processors employ a `difference over sum' signal processing technique~\cite{Apsimon15} as follows. The signals from the top and bottom strips are added using a resistive coupler, and subtracted using a 180$^\circ$ hybrid. An external, continuous, machine-derived local oscillator (LO) signal is used to down-mix the radio-frequency (RF) sum and difference signals to produce the baseband signals $V_\Sigma$ and $V_\Delta$, respectively. These signals can then be digitized, and the beam position is calculated from the ratio $\frac{V_\Delta}{V_\Sigma}$.

The stripline BPMs have a demonstrated position resolution of $291 \pm 10$~nm at a bunch charge of $\sim 1$~nC, with a linear response range of $\pm 500$~$\mu$m~\cite{Apsimon15}. The scaling of the BPM resolution with the inverse of the bunch charge, down to a charge of 0.3~nC, has been demonstrated~\cite{Apsimon15}. Hence, with a factor of three signal attenuation, the ILC dynamic range (Sec.~\ref{sec:Intro}) is achieved with a position resolution of $\sim 1$~$\mu$m. Further attenuation would be used to compensate for the higher bunch charge at the ILC, whose design value is $\sim 3$~nC (Table~\ref{tab:ILCParameters}). The signal processor latency has been measured to be $15.6 \pm 0.1$~ns \cite{Apsimon15}.

\subsection{\label{sec:FONT5ABoard}FONT5A digital feedback board}

\begin{figure}
  \centering
  \includegraphics[width=\columnwidth]{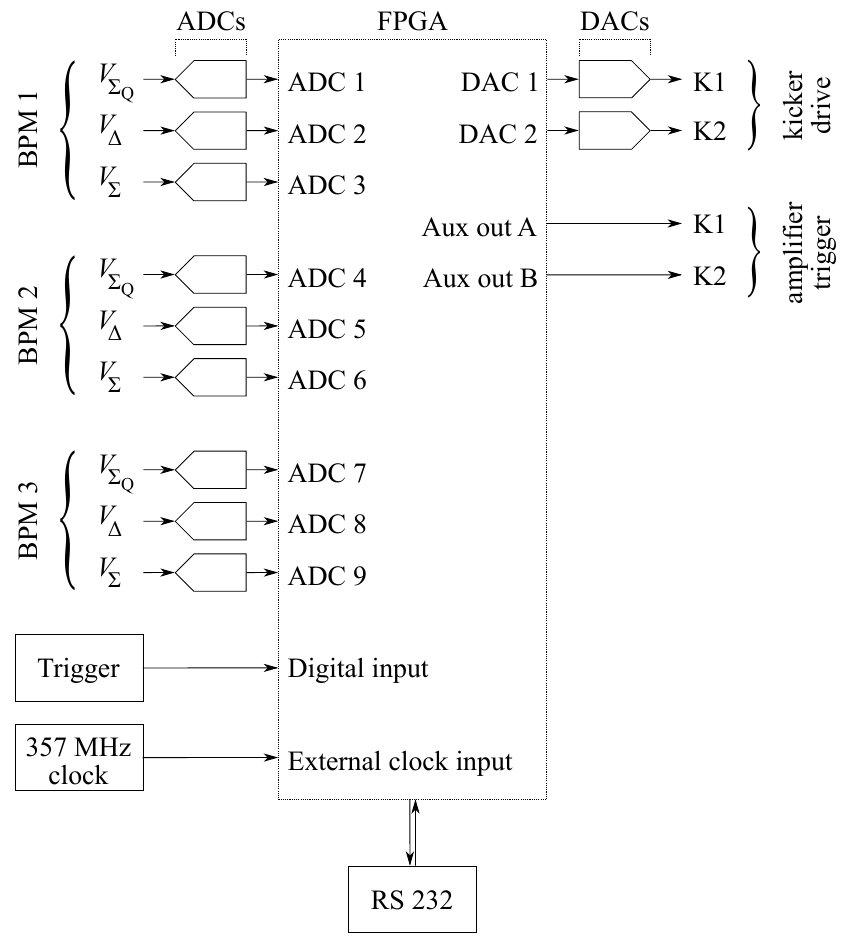}
	\caption{Schematic of the FONT5A digital feedback board.}
	\label{fig:FONT5ABoard}
\end{figure}

The stripline BPM signal processor outputs are digitized in the FONT5A digital feedback board (Fig.~\ref{fig:FONT5ABoard}). The board consists of a Xilinx Virtex-5 XC5VLX50T field programmable gate array (FPGA) \cite{Xilinx}, nine Texas Instruments ADS5474 14-bit analogue-to-digital converters (ADCs) \cite{ADC} and two Analog Devices AD9744 14-bit digital-to-analogue converters (DACs)~\cite{DAC} whose output is used to drive the kicker amplifier (Fig.~\ref{fig:FeedbackSchematic}). An external trigger, preceding the extraction of the bunches from the ATF damping ring (Fig.~\ref{fig:ATFSchematic}), is used to synchronize feedback operation to the bunch arrival time, as well as to control the timing of the digitization of the BPM signals. Each ADC is clocked with a 357~MHz signal synchronized with the ATF damping ring RF master oscillator. The feedback calculation runs on the FPGA, but data are also sent serially from the board via a universal asynchronous receiver/transmitter (UART) over RS232 to a local computer for offline storage and data analysis.


For a given train of bunches, the FONT5A board effectively measures the position of the first bunch and attempts to zero the position of subsequent bunches. For an ideal feedback system, the position $Y_2$ of the second bunch with feedback on is:

\begin{equation}
\label{eq:CorrectedBunch2Position}
Y_2 = y_2 - g y_1 - \delta_2,
\end{equation}

\noindent where $y_1$ and $y_2$ are the incoming, uncorrected positions of the first and second bunches, respectively, $g$ is the feedback gain and $\delta_2$ is a constant offset applied to account for the static position offset between the first and second bunches. The gain is set so the offset of the beam position from zero is measured at the first bunch, and is used to fully correct the position of the second, and thus $g \approx 1$. For subsequent bunches ($n \ge 3$):

\begin{equation}
\label{eq:CorrectedBunchnPosition}
Y_n = y_n - g y_1 - g \sum_{i=2}^{n-1} Y_i - \delta_n,
\end{equation}

\noindent where $y_n$ and $Y_n$ are the uncorrected and corrected positions of the $n^\textrm{th}$ bunch, respectively, and $\delta_n$ is a constant offset. The corrections applied to previous bunches are accumulated in the delay loop register on the FPGA, constituting the memory of the total correction performed so far to the bunches in the train. Thus, the delay loop maintains the corrected position for all subsequent bunches.  






Operating the feedback in single-loop mode, the K2 drive signal issued by the DAC, $V_\textrm{DAC}$, is calculated by:

\begin{equation}
\label{eq:KickCalculation}
V_\textrm{DAC} = g K \frac{V_\Delta}{V_\Sigma} + D,
\end{equation}

\noindent where $K$ is the kick factor, $V_\Delta$ and $V_\Sigma$ are the digitized difference and sum signals from the P3 signal processor, and $D$ is the value stored in the delay loop. The kick factor can be calculated from the slope, $H$, of the measured beam position $\frac{V_\Delta}{V_\Sigma}$ versus set values of the K2 drive $V_\textrm{DAC}$:

\begin{equation}
\label{eq:GainCalculation}
K = - \frac{V_\textrm{DAC}}{(\frac{V_\Delta}{V_\Sigma})} = - \frac{1}{H}
\end{equation}

\noindent where the minus sign originates from the requirement that the feedback subtracts the measured offset so as to zero the beam position.

\begin{figure*}
	\centering
	\setlength\fboxsep{0pt}
	\includegraphics[width=\textwidth]{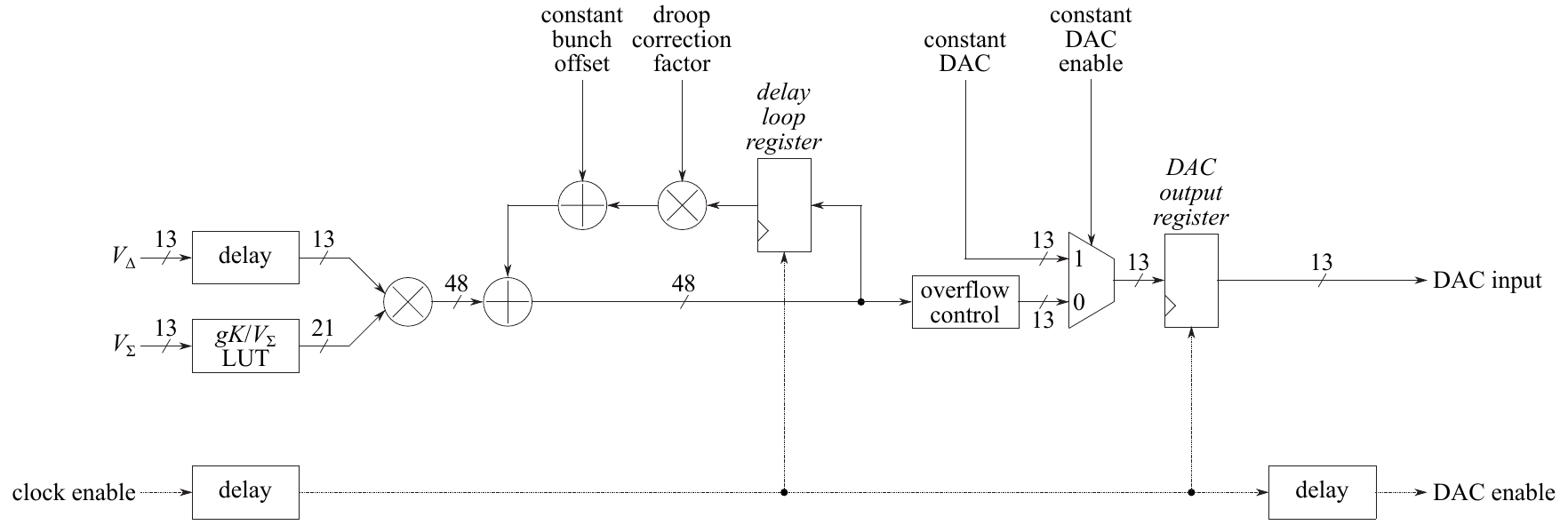}
	\caption{Simplified representation of the feedback signal processing architecture on the FPGA. The signal bus lines are annotated to show the number of bits represented at each stage of the processing. Note that, for simplicity, clock signals are not shown in the diagram, and hence the schematic does not represent the actual timing of the feedback processing.}
	\label{fig:FirmwareSchematic}
\end{figure*}

The FONT5A board firmware implementation is shown in Fig.~\ref{fig:FirmwareSchematic}. A look-up table (LUT) is implemented in core memory resources on the FPGA, and is used to obtain the product of $gK$ and the reciprocal of the incoming $V_\Sigma$ signal whilst the $V_\Delta$ signal is delayed accordingly. The two signals are then  multiplied together, before entering both the DAC and delay loop. The timing is set such that only the signals calculated from the sampled bunches are strobed onto the delay loop and DAC output registers. The value stored in the delay loop can be multiplied by a droop correction factor to compensate for the effective roll-off at low frequencies, due to the transformer coupling between the ADC and the kicker amplifier, in the output signal, and consequent droop in the step response. A constant bunch offset term can also be added in the delay loop to correct for any static offset between the positions of consecutive bunches. The value stored in the delay loop is then added to the $V_\Delta \times \frac{gK}{V_\Sigma}$ value measured for each bunch in turn. The relevant 13 bits are selected to constitute the DAC output; if the calculation has overflowed the 13 bit bound, the resultant 13-bit value is saturated at its minimum or maximum value. The DAC output can be set to a constant DAC value, and this feature is used for calibrating the effect of the kicker on the beam position.

\subsection{\label{sec:Kicker}Kicker and amplifier}

\begin{figure}
  \centering
  \includegraphics[width=\columnwidth]{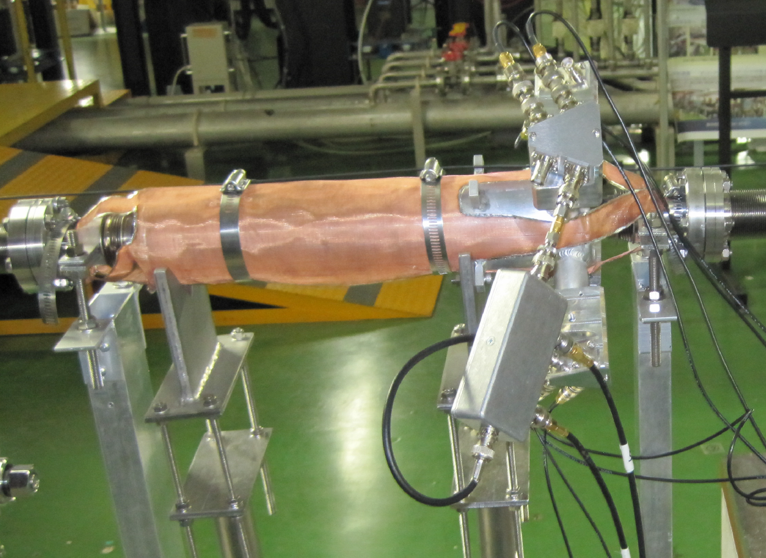}
	\caption{Photograph of the kicker K2 in the ATF beamline.}
	\label{fig:Kicker}
\end{figure}

\begin{figure*}
	\centering
	\setlength\fboxsep{0pt}
	\includegraphics[width=\textwidth]{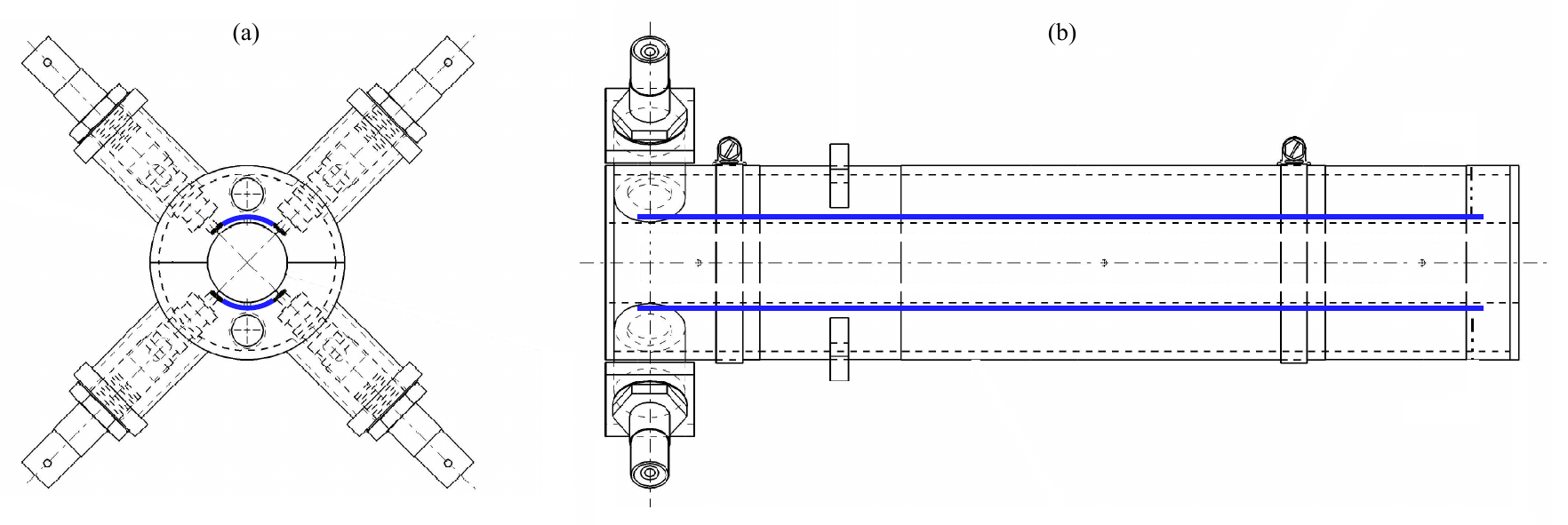}
	\caption{(a) Transverse and (b) longitudinal views of the stripline kicker used in the FONT system \cite{JollyDPhil}. The two conducting strips are shown in blue: they are linked together at their right hand ends in (b).}
	\label{fig:KickerDrawing}
\end{figure*}

The kickers (Fig.~\ref{fig:Kicker}), provided by the SLAC laboratory, each consist of two parallel conducting strips, approximately 30~cm in length, placed along the top and bottom of a ceramic section of beampipe, as shown in the technical drawing in Fig.~\ref{fig:KickerDrawing}. By being driven with input signals at the downstream end and with the electrodes shorted together at the upstream end, the kicker deflects the beam in the vertical plane. The kicker drive signal from the FONT5A board, with a maximum range of $\pm 2$~V, is amplified with a custom-built amplifier, which delivers a high current with a fast rise time. The required amplifier was developed and manufactured for this purpose by TMD Technologies Ltd \cite{TMD} and can provide up to $\pm 30$~A of drive current with a rise time of 35~ns from the time of the drive signal arrival to that of 90\% of peak output. The output pulse length is specified to be up to 10~$\mu$s. The amplifier needs to be triggered in advance of the bunch arrival; the trigger signal is generated by the FONT5A board (Fig.~\ref{fig:FONT5ABoard}).


\section{\label{sec:Results}System performance}

\subsection{\label{sec:Latency}System latency}

The system latency was designed to be lower than the inter-bunch spacing. The latency was hence measured conveniently by systematically adding controlled extra delay until the feedback correction signal arrived too late to affect the beam. In practice this was performed by enabling a constant DAC output and then delaying it (Fig.~\ref{fig:FirmwareSchematic}). An effective bunch spacing can be defined as the sum of the actual bunch spacing and the added delay.

\begin{figure}
  \centering
  \includegraphics[width=\columnwidth]{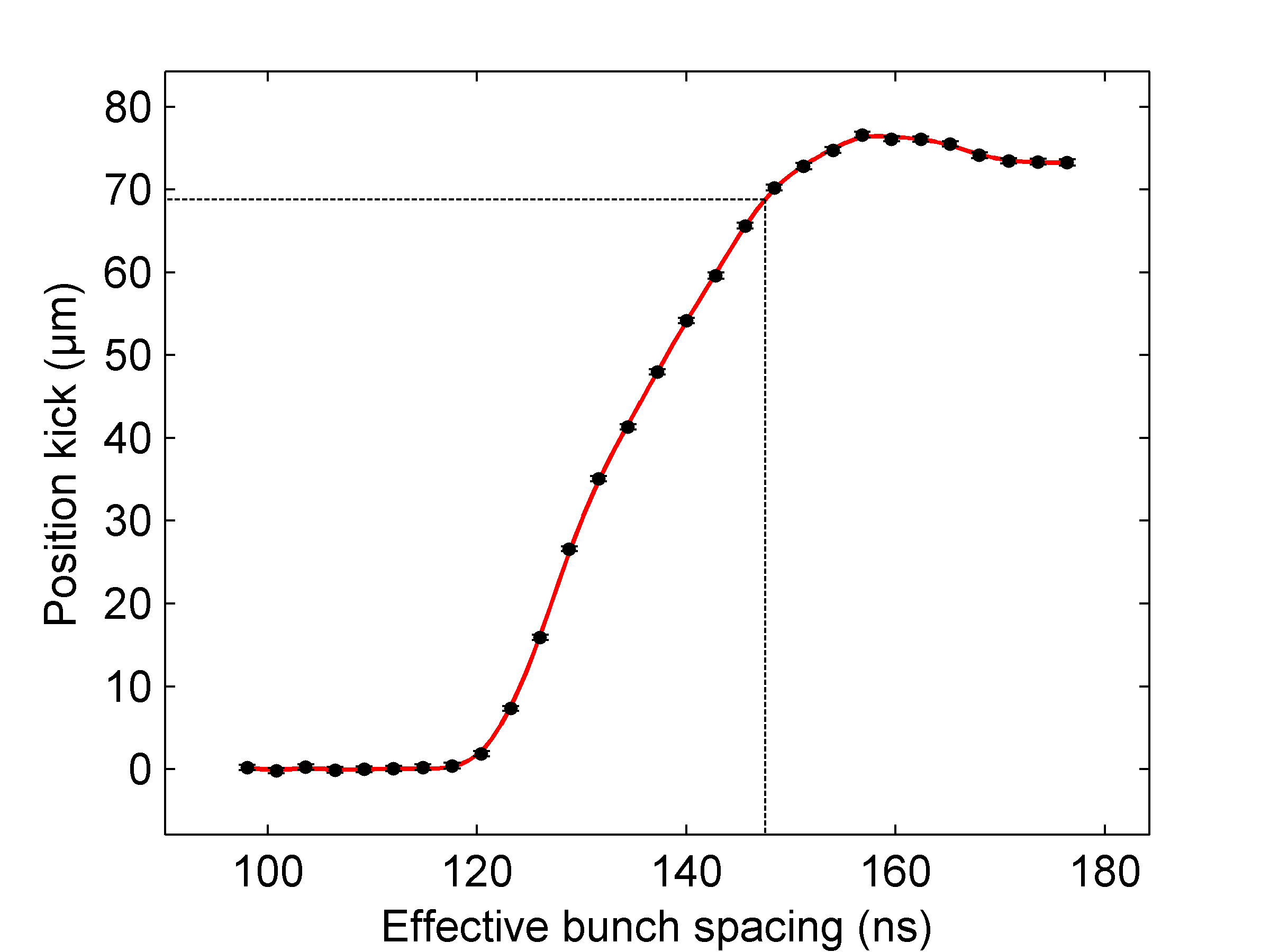}
	\caption{Average difference between the kicked and unkicked positions of bunch 2 versus bunch spacing, for a constant kick corresponding to a DAC setting of 2000 counts. The errors (calculated as the sum in quadrature of the errors on the mean kicked and unkicked positions) are given. The red line is a cubic spline fit to the data and the black lines indicate the point at which 90\% of the full kick is delivered.}
	\label{fig:LatencyScan}
\end{figure}

Data were taken as a function of added delay with interleaved kicked and unkicked beam to mitigate against beam drift, and averaged at each setting to remove the effect of beam jitter on the measurement. Fig.~\ref{fig:LatencyScan} shows the average difference between kicked and unkicked beam position versus effective bunch spacing. The system latency is defined as the point at which 90\% of the full kick is delivered, and yields a latency of 148~ns (Fig.~\ref{fig:LatencyScan}).


\subsection{\label{sec:KickerScan}Kicker linearity and range}

\begin{figure}
  \centering
  \includegraphics[width=\columnwidth]{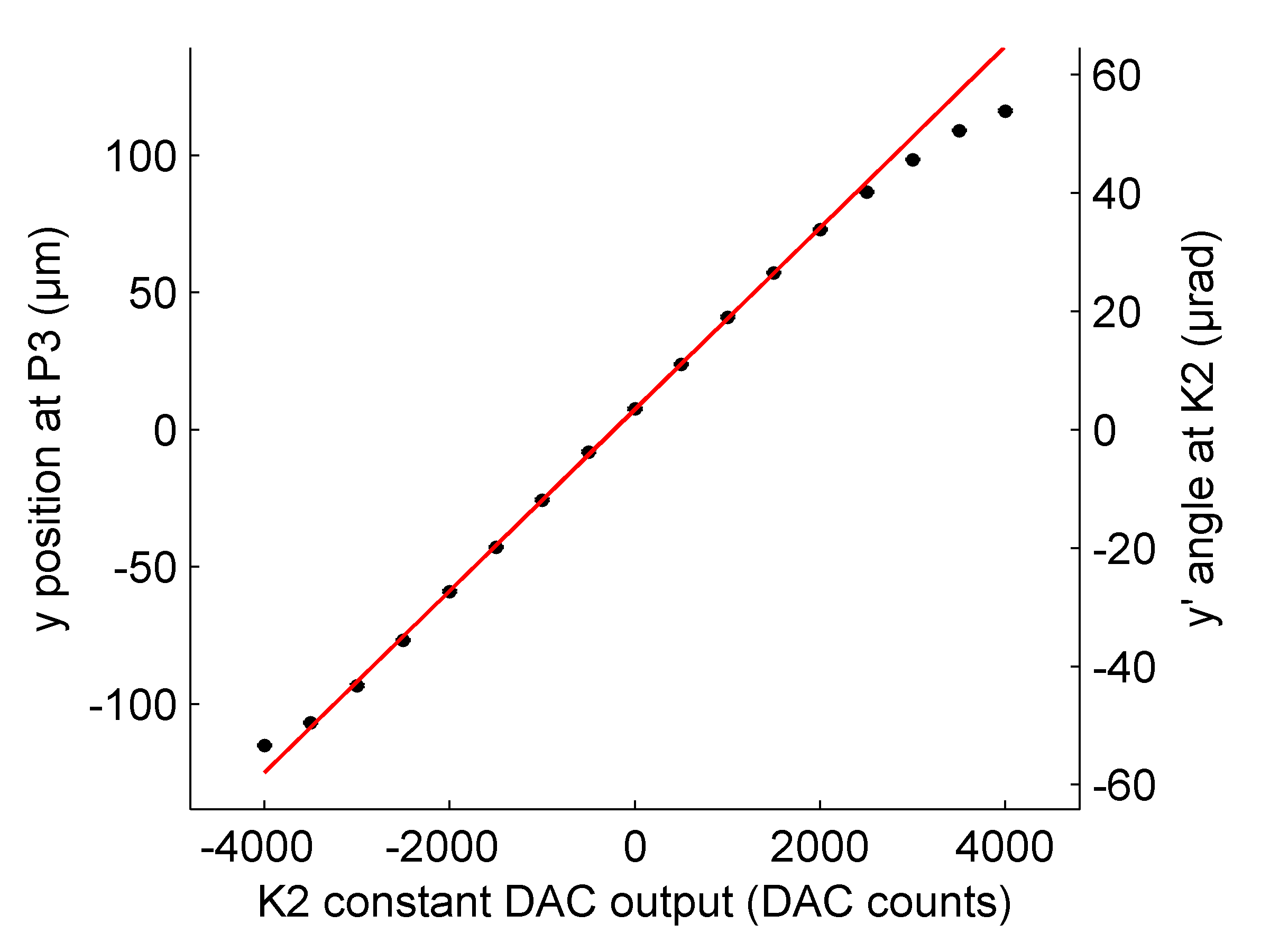}
	\caption{Vertical beam position at P3 (left-hand scale) versus constant kick applied at K2. The right-hand scale shows the corresponding $y'$ kick. The errors on the mean positions are given. The red line is a linear $\chi^2$ fit to the central nine data points.}
	\label{fig:KickerScan}
\end{figure}

The amplifier and kicker performance were tested by systematically varying the amplifier drive signal and measuring the beam displacement at P3 (Fig.~\ref{fig:KickerScan}). The angular kick imparted to the beam by K2, $y'_\textrm{K2}$, can be reconstructed from the measured displacement at P3, $y_\textrm{P3}$, using element $M_{34}$ of the $6 \times 6$ linear beam transfer matrix $M$ between K2 and P3; this was calculated  by using the MAD~\cite{MAD} model of the ATF2 beamline.

\begin{equation}
\label{eq:M34}
y_\textrm{P3} = M_{34} y'_\textrm{K2}.
\end{equation}

\noindent From Fig.~\ref{fig:KickerScan} it can be seen that a linear kicker response is observed over a correction range of $\pm 75$~$\mu$m, corresponding to a kick range of $\sim \pm 35$~$\mu$rad provided by K2. This scales to $\sim \pm 180$~nrad for the 250~GeV ILC beam energy, which exceeds the requirements discussed in Sec.~\ref{sec:Intro}.

\subsection{\label{sec:FeedbackResults}Feedback results}

The ATF was configured so as to deliver successive trains of three bunches with a bunch separation of 154~ns. For the subsequent measurements the beam was approximately centered vertically in MQF15X using an upstream corrector, and then centered in P3 using the BPM mover. The feedback was operated in interleaved mode, whereby alternate trains were subjected to feedback off and on. The data with feedback off were used to characterize the incoming beam and to track drifts in the beam conditions. The feedback system was operated with $g = 1$ as the bunches have similar position jitters (Table~\ref{tab:MeasuredFeedbackP3}) and the positions of the bunches are highly correlated (Table~\ref{tab:BunchToBunchCorrelationP3}).

\begin{figure}
  \centering
  \includegraphics[width=\columnwidth]{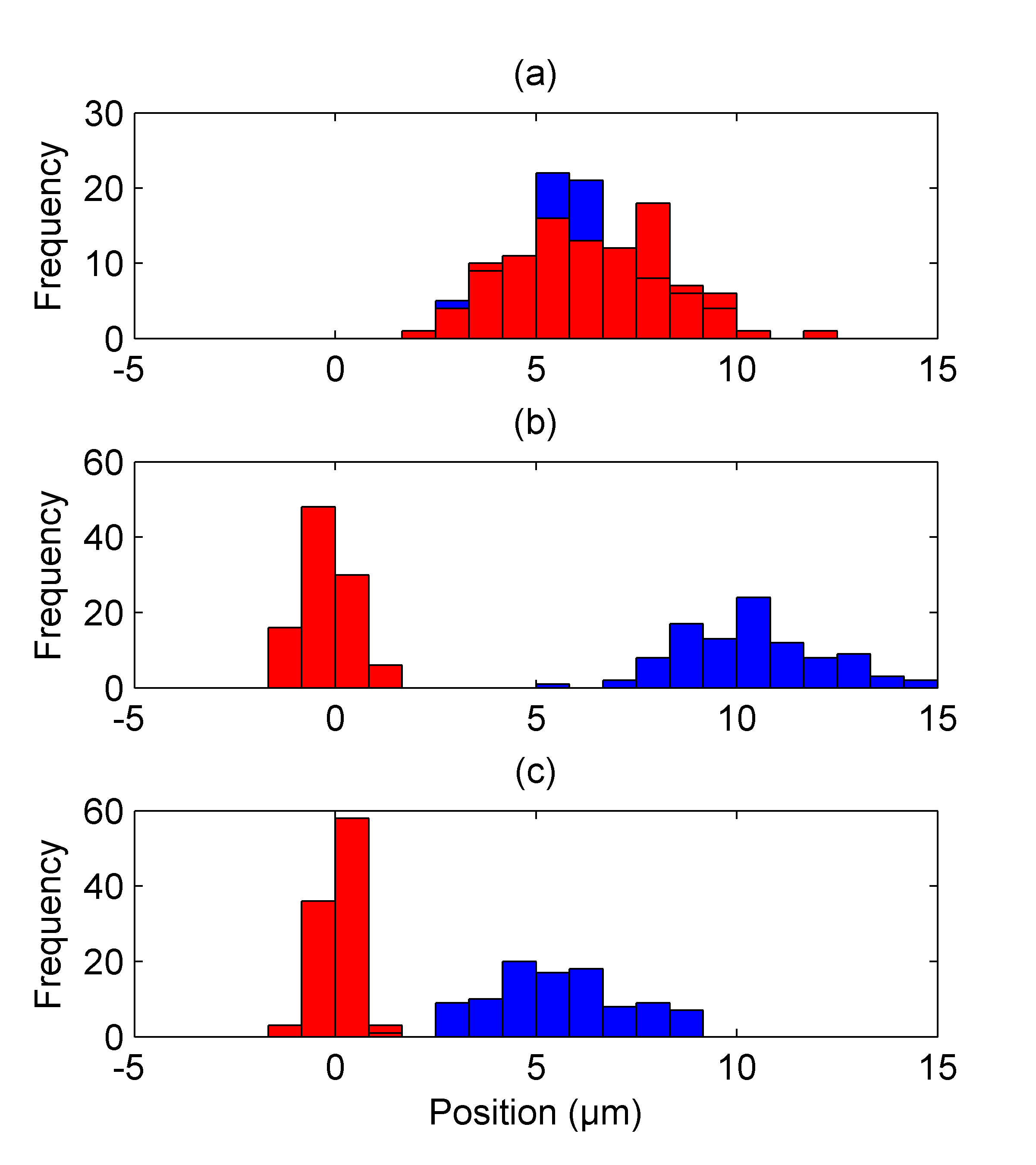}
	\caption{Distribution of beam positions measured at P3 with feedback off (blue) and on (red) for (a) the first, (b) the second, and (c) the third bunch in each train.}
	\label{fig:MeasuredFeedback}
\end{figure}

The beam position recorded at P3 is shown in Fig.~\ref{fig:MeasuredFeedback}, for a data set with 100 trains with feedback on interleaved with 100 trains with feedback off. The first bunch in each train is not affected by the feedback as this bunch is only measured, but not corrected. The second and third bunches show the effect of the feedback: the corrected beam positions are centered on zero and the spread of beam positions is reduced.

The mean beam position and the beam jitter, defined to be the standard deviation $\sigma$ of the position distribution, are listed in Table~\ref{tab:MeasuredFeedbackP3}. For a distribution with $n$ triggers, the standard error $\sigma_\mu$ on the mean position is given by $\sigma_\mu = \frac{\sigma}{\sqrt{n}}$ and the standard error $\sigma_\sigma$ on the jitter is given by $\sigma_\sigma = \frac{\sigma}{\sqrt{2n}}$ \cite{PDG18}.

The feedback acts to reduce the incoming beam jitter from $\sim 1.6$~$\mu$m to $0.45$~$\mu$m. The incoming bunch train is observed (Fig.~\ref{fig:MeasuredFeedback}) to have a static bunch-to-bunch position offset, whereby bunch 2 lies roughly 5~$\mu$m higher than bunches 1 and 3. By making use of the constant offset $\delta_n$ introduced in Eq.~\ref{eq:CorrectedBunchnPosition}, the feedback centers the mean position of bunches two and three to within $0.25$~$\mu$m of P3's electrical center.

\begin{table}
\caption{Mean beam position and beam jitter measured at P3 with feedback off and on for bunches 1, 2 and 3. Standard errors are given for both the mean positions and jitters.}
\label{tab:MeasuredFeedbackP3}
\centering
\begin{tabular}{ccccc}
																																																				 \hline \hline 
					& \multicolumn{2}{c}{Mean position ($\mu$m)} & \multicolumn{2}{c}{Position jitter ($\mu$m)} \\               
    Bunch & Feedback off     & Feedback on             & Feedback off    & Feedback on                \\ \hline        
		1     & $5.97 \pm 0.16$  & $6.37 \pm 0.19$         & $1.61 \pm 0.12$ & $1.93 \pm 0.14$            \\
		2     & $10.39 \pm 0.17$ & $-0.25 \pm 0.06$        & $1.65 \pm 0.12$ & $0.60 \pm 0.04$            \\
		3     & $5.58 \pm 0.17$  & $0.04 \pm 0.05$         & $1.63 \pm 0.12$ & $0.45 \pm 0.03$            \\ \hline \hline 
\end{tabular}
\end{table}

An incoming bunch-to-bunch position correlation in excess of $94\%$ was measured for this data set (Table~\ref{tab:BunchToBunchCorrelationP3}). A high correlation is required in order to obtain a substantial reduction in position jitter. The feedback acts to remove the correlated position components between the bunches, and was able to reduce the correlation to almost zero.

\begin{table}
\caption{Bunch-to-bunch correlation with feedback off and on at P3. One standard deviation confidence intervals are given.}
\label{tab:BunchToBunchCorrelationP3}
\centering
\begin{tabular}{ccc}
																																 \hline \hline 
											 & \multicolumn{2}{c}{Correlation (\%)}	\\ 							 
											 & Feedback off & Feedback on	 					\\ \hline				 
		Bunch 1 to bunch 2 & $+94 \pm 1$	& $-11 \pm 10$ 					\\
		Bunch 2 to bunch 3 & $+96 \pm 1$	& $-28 \pm 9$  					\\ \hline \hline 
\end{tabular}
\end{table}

The expected feedback performance can be estimated by taking the standard deviation of the terms in the feedback algorithm defined in Eq.~\ref{eq:CorrectedBunch2Position}, given $g=1$:

\begin{equation}
\label{eq:PredictedFeedbackPerformance}
\sigma_{Y_n}^2 = \sigma_{y_n}^2 + \sigma_{y_{n-1}}^2 - 2 \sigma_{y_n} \sigma_{y_{n-1}} \rho_{y_n y_{n-1}},
\end{equation}

\noindent where $\sigma_{Y_n}$, $\sigma_{y_n}$ and $\sigma_{y_{n-1}}$ are the respective standard deviations of the distributions of $Y_n$, $y_n$ and $y_{n-1}$ and $\rho_{y_n y_{n-1}}$ is the correlation between $y_n$ and $y_{n-1}$. Substituting the respective measured values (Tables~\ref{tab:MeasuredFeedbackP3} and \ref{tab:BunchToBunchCorrelationP3}) into Eq.~\ref{eq:PredictedFeedbackPerformance} yields predicted corrected jitters of $\sigma_{Y_2} = 0.58$~$\mu$m and $\sigma_{Y_3} = 0.45$~$\mu$m. These values agree with the measured jitters of $0.60 \pm 0.04$~$\mu$m and $0.45 \pm 0.03$~$\mu$m, respectively, which indicates that the feedback performed optimally. 



\begin{figure}
	\centering
  \includegraphics[width=\columnwidth]{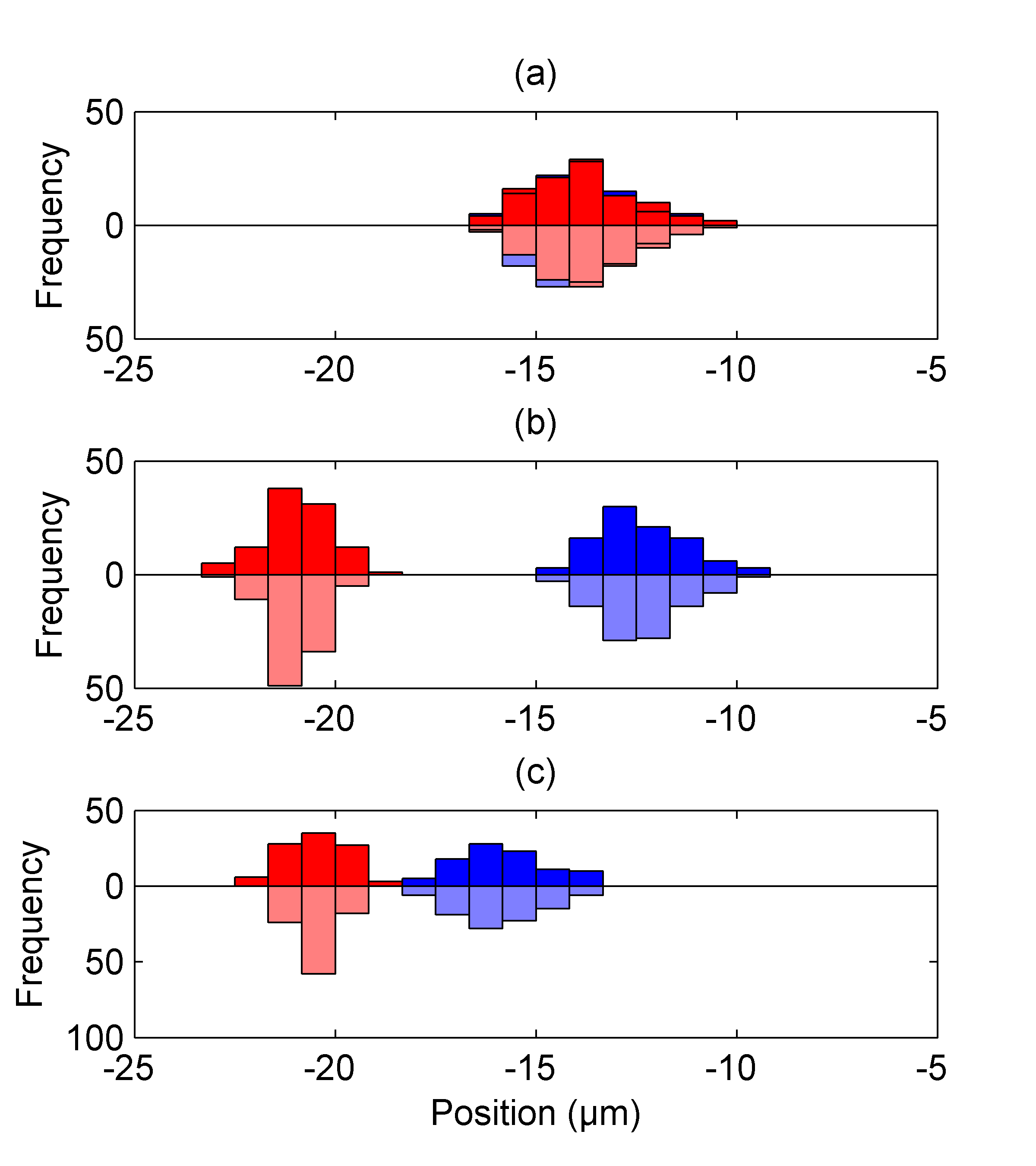}
	\caption{Distributions of beam positions at MQF15X with feedback off (blue) and on (red) for (a) the first, (b) the second, and (c) the third bunches. The darker, positive bars show the measured positions; the lighter, negative bars show the positions measured at P2 and P3 propagated to MQF15X. The mean propagated positions have been adjusted to match those  of the respective measurements, as the beam propagation model does not take the actual BPM positions into account.}
	\label{fig:PropagatedFeedback}
\end{figure}

The corresponding beam position measurements recorded in the downstream witness BPM MQF15X are shown in Fig.~\ref{fig:PropagatedFeedback}. A substantial reduction in jitter is apparent. The feedback performance agrees well with that expected from propagating the measured beam positions at BPMs P2 and P3 (Fig.~\ref{fig:EXTFFSchematic}) using linear transfer matrices calculated from the ATF MAD model. 


\begin{figure*}
  \centering
	\begin{minipage}{\columnwidth}
		\centering
		\includegraphics[width=\columnwidth]{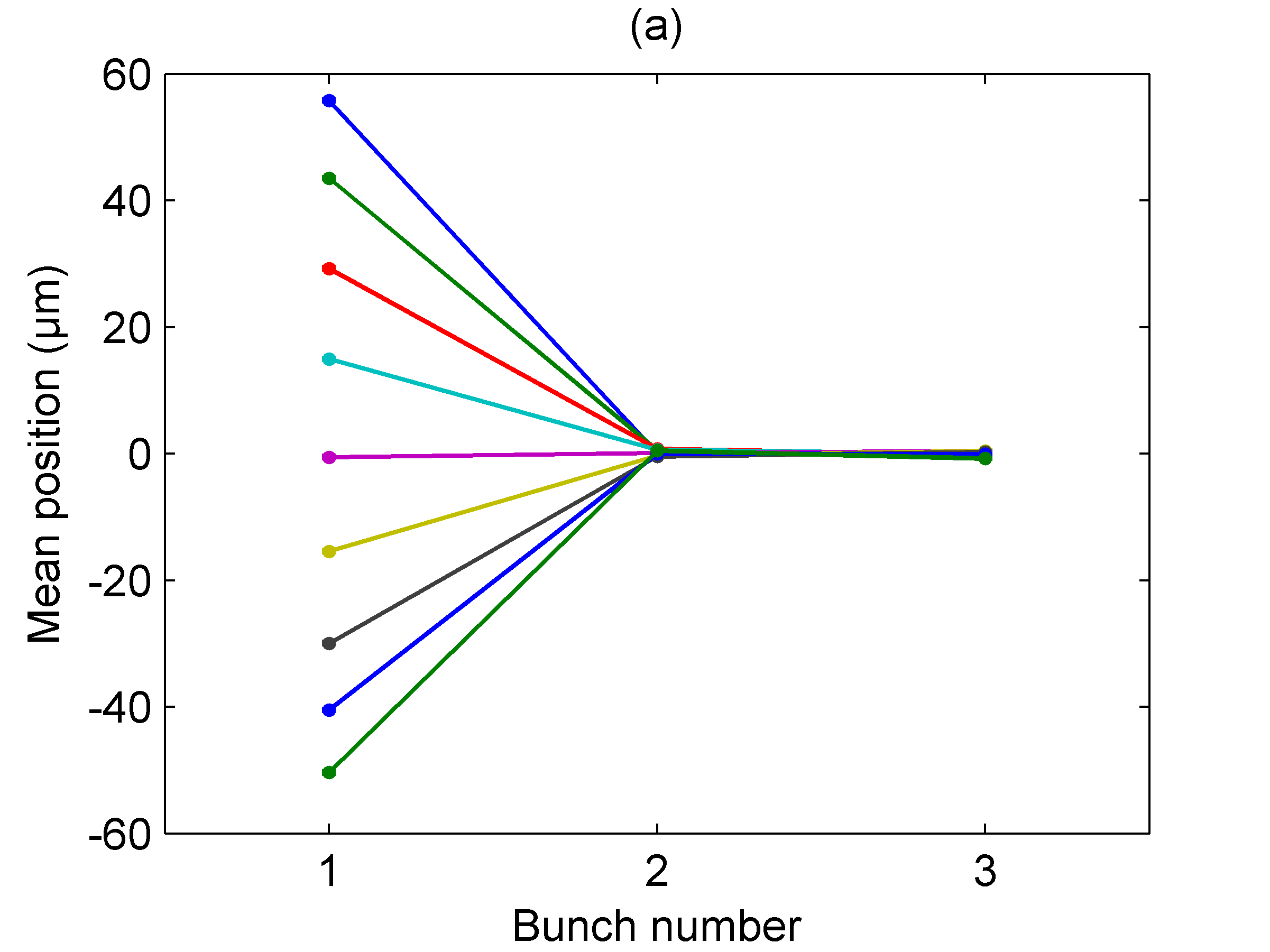}
	\end{minipage}
	\hfill 
	\begin{minipage}{\columnwidth}
		\centering
		\includegraphics[width=\columnwidth]{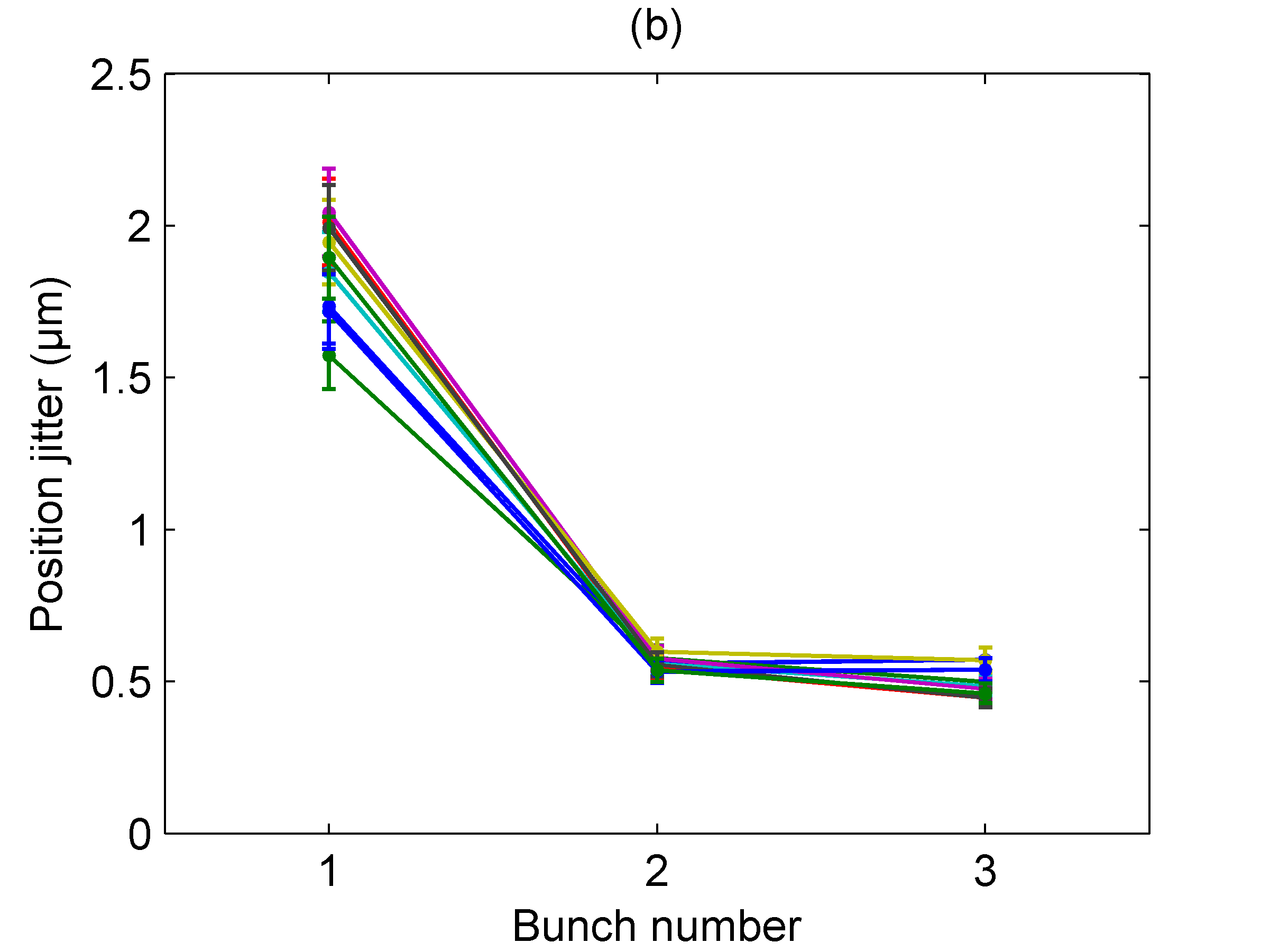}
	\end{minipage}
	\caption{(a) Mean position and (b) position jitter measured at P3 with feedback on versus bunch number for nine incoming beam orbit settings (color coded). Standard errors are given.}
	\label{fig:TShirtPlot}
\end{figure*}

In order to assess the feedback operation over a wide correction range, the vertical position of the beam arriving at P3 was swept through a range of approximately $\pm 60$~$\mu$m by varying a corrector magnet located upstream of K2. The results (Fig.~\ref{fig:TShirtPlot}) show that the mean positions of the second and third bunches are zeroed and the spread of positions is consistently reduced to around 500~nm.



\begin{figure*}
	\centering
	\setlength\fboxsep{0pt}
	\includegraphics[width=\textwidth]{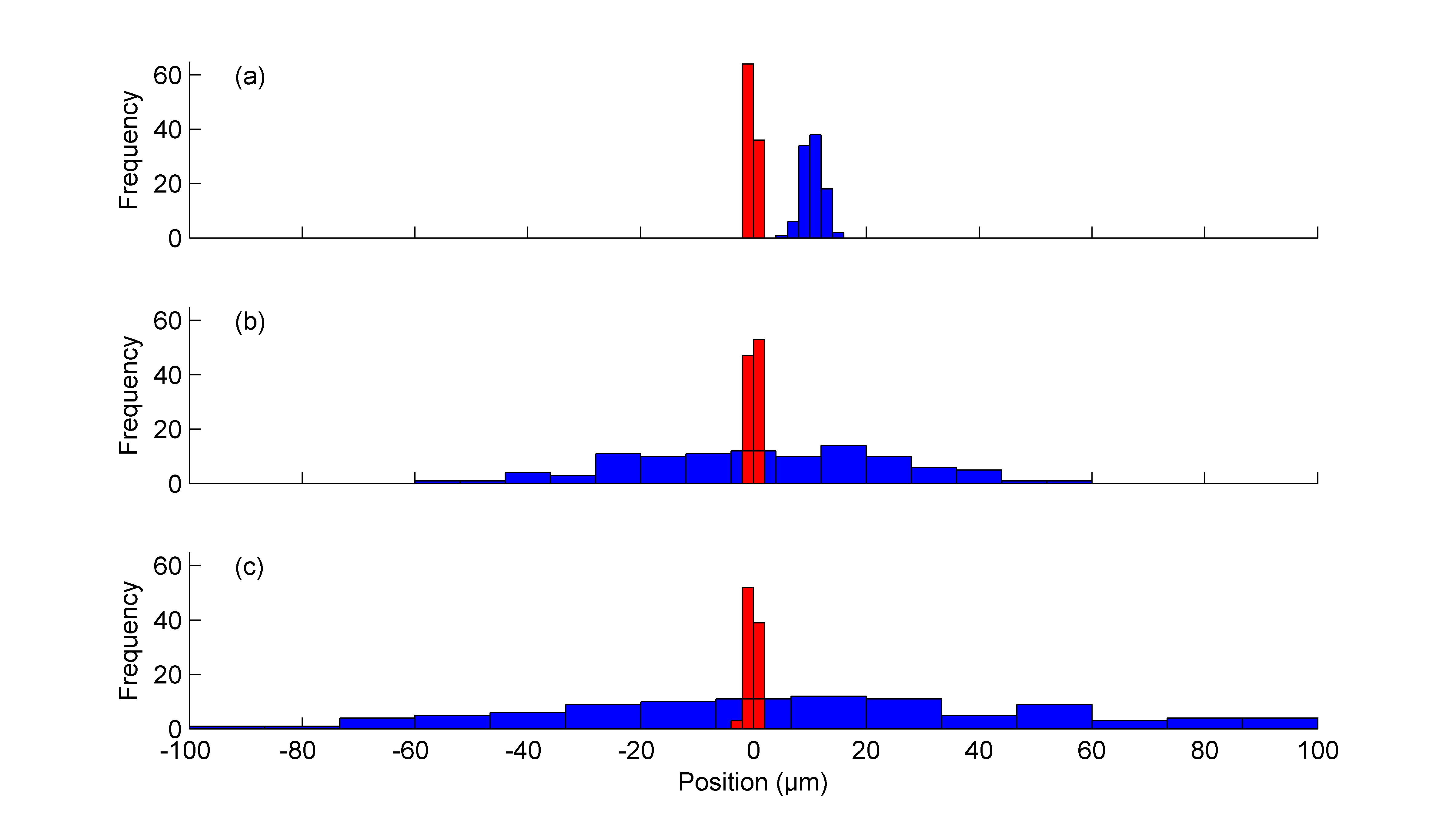}
	\caption{Distributions of positions with feedback off (blue) and feedback on (red) for bunch 2 at P3 with incoming, uncorrected position jitters of (a) $\sim 2$~$\mu$m,  (b) $\sim 22$~$\mu$m and (c) $\sim 45$~$\mu$m.}
	\label{fig:RandomJitterBunch2}
\end{figure*}

As an additional test, two vertical steering magnets were used to enhance the incoming beam jitter. The magnets were set up so as to apply a random kick conforming to a pre-defined distribution with the kick updated successively at the train repetition frequency. The feedback was observed (Fig.~\ref{fig:RandomJitterBunch2}) to successfully center and stabilize the beam, even when the full spread of uncorrected positions reaches $\pm 100$~$\mu$m.

The ability of the feedback system to stabilize the beam at the feedback BPM to $\sim 0.5$~$\mu$m implies that such a system could be used to stabilize the outgoing beam from the ILC IP to $\sim 0.5$~$\mu$m as measured at the ILC IP feedback BPM; that is, to stabilize the outgoing beam deflection angle to $\sim 0.12$~$\mu$rad. From Fig.~\ref{fig:LuminosityVsAngle}, this is equivalent to a luminosity stabilization to within 0.1\% of the nominal value.


\section{\label{sec:Conc}Conclusions}

An intra-train position feedback system has been designed to achieve and maintain collisions at the ILC, and a prototype has been developed, commissioned and tested at the ATF. The beam position is measured using a stripline BPM with analogue signal-processing electronics. The outputs are processed on an FPGA-based digital board used to calculate and deliver a correction signal, which is amplified by a high-current drive amplifier and applied to a stripline kicker. All components have been designed for minimum latency, with an overall feedback latency of 148~ns, allowing bunch-to-bunch feedback at the ILC. The stripline BPM has a position resolution of $291 \pm 10$~nm and a linear range of $\pm 500$~$\mu$m and satisfies the ILC requirements. The kicker response is linear over a correction range of over $\pm 60$~$\mu$m measured at the feedback BPM which satisfies the ILC requirements. The feedback system has been used to successfully stabilize the second and third bunches in a three-bunch train with 154~ns bunch spacing, where the first bunch is used as a pilot bunch. The propagation of the correction has been verified by using an independent stripline BPM located downstream of the feedback system. The performance is maintained on sweeping the incoming beam orbit through $\pm 50$~$\mu$m or enhancing the spread of incoming beam orbits by up to $\pm 100$~$\mu$m, which exceeds the equivalent ILC operating range. A comparison of the performances demonstrated here with those required for the ILC is given in Table~\ref{tab:ILCATFPerformanceComparison}. The system has been demonstrated to meet the BPM resolution, beam kick and latency requirements for the ILC.



\begin{table}
\caption{Comparison of the IP feedback performance required at the ILC with that achieved by the FONT feedback system at ATF.}
\label{tab:ILCATFPerformanceComparison}
\centering
\begin{tabular}{cccc}
																																																					 \hline \hline 
											          &        & ILC           & ATF																					\\ \hline				 
		Energy per beam             & GeV    & 250           & 1.3																					\\
		IP feedback latency         & ns     & 554           & 148																					\\
		BPM dynamic range           & $\mu$m & $\pm 1400$    & $\pm 1500$																		\\
		BPM resolution              & $\mu$m & $\sim 1$      & $\sim 1$																			\\
		Beam angle correction range & nrad   & $\sim \pm 60$ & $\sim \pm 180$\textsuperscript{\textdagger}	\\ \hline \hline 
		\multicolumn{4}{c}{\textsuperscript{\textdagger} scaled by the ATF/ILC beam energy ratio}
\end{tabular}
\end{table}


Having built a prototype system which meets the technical requirements for the ILC, the next step is to implement the demonstrated performance in a simulation of the beam collision feedback system and evaluate its luminosity recovery capability subject to realistic beam imperfections. This requires detailed modelling of beam transport through the ILC beamline complex, from the exit of the damping rings through to the interaction region, and must incorporate expected beam imperfections including those due to static component misalignments as well as dynamic misalignments resulting from ground motions, facilities noise, and the performance of upstream beam feedback and feed-forward systems. Some earlier studies have been performed \cite{White06, Resta08, Resta09, Bodenstein17}, and a significant update is in progress using the latest ILC design and the collision feedback system performance reported here; this is the subject of a paper in preparation.


\section*{Acknowledgments}
We thank the KEK ATF staff for their outstanding logistical support and providing the beam-time and the necessary stable operating conditions for this research. In addition we thank our ATF2 collaborators for their help and support; in particular the University of Valencia IFIC group for providing the BPM mover system. We acknowledge financial support for this research from the UK Science and Technology Facilities Council via the John Adams Institute, University of Oxford, and CERN, CLIC-UK Collaboration, Contract No.~K E1869/DG/CLIC. The research leading to these results has received funding from the European Commission under the Horizon 2020 / Marie Sk\l{}odowska-Curie Research and Innovation Staff Exchange (RISE) project E-JADE, Grant agreement No.~645479.

\bibliography{ILCFBpaperReferences}

\end{document}